\documentclass[aps,prl,twocolumn,superscriptaddress]{revtex4}
\usepackage{graphicx}
\usepackage{amssymb}
\usepackage{bm}

\usepackage{graphpap}
\begin{document}

\title{First principle theory of correlated transport through nano-junctions}
\author{A. \surname{Ferretti}}
\affiliation{INFM National Center on nanoStructures
             and bioSystems at Surfaces ($S^3$)
             and Dipartimento di Fisica, Universit\`a di
             Modena e Reggio Emilia, 41100 Modena, Italy}
\author{A. \surname{Calzolari}}
\affiliation{INFM National Center on nanoStructures
             and bioSystems at Surfaces ($S^3$)
             and Dipartimento di Fisica, Universit\`a di
             Modena e Reggio Emilia, 41100 Modena, Italy}
\author{R.   \surname{Di Felice}}
\affiliation{INFM National Center on nanoStructures
             and bioSystems at Surfaces ($S^3$)
             and Dipartimento di Fisica, Universit\`a di
             Modena e Reggio Emilia, 41100 Modena, Italy}
\author{F. \surname{Manghi}}
\affiliation{INFM National Center on nanoStructures
             and bioSystems at Surfaces ($S^3$)
             and Dipartimento di Fisica, Universit\`a di
             Modena e Reggio Emilia, 41100 Modena, Italy}
\author{M. J. \surname{Caldas}}
\affiliation{Instituto de F{\'\i}sica, Universidade de S{\~a}o
             Paulo,  05508-900 S{\~a}o Paulo, Brazil}
\author{M. \surname{Buongiorno Nardelli}}
\affiliation{Department of Physics, North Carolina State
             University, Raleigh, NC 27695, USA}
\affiliation{CCS-CSM, Oak Ridge National Laboratory, Oak Ridge,
             TN 37831, USA }
\author{E.  \surname{Molinari}}
\affiliation{INFM National Center on nanoStructures
             and bioSystems at Surfaces ($S^3$)
             and Dipartimento di Fisica, Universit\`a di
             Modena e Reggio Emilia, 41100 Modena, Italy}
\date{\today}

\begin{abstract}
\noindent
We report the inclusion of electron-electron correlation in the
calculation of transport properties within an {\it ab initio}
scheme. A key step is the reformulation of Landauer's approach in
terms of an effective transmittance for the interacting electron
system. We apply this framework to analyze the effect of short range
interactions on Pt atomic wires and discuss the coherent and
incoherent correction to the mean-field approach.
\end{abstract}

\maketitle

One of the most pressing problems in nanotechnology is the need
for recasting all the know-how about mesoscopic transport physics
into the fully quantum mechanical limit appropriate for atomic
scale phenomena. In the case of electronic and transport
properties of atomic and molecular conductors,
we must address at the microscopic level both the chemical complexity of
the conductor and the complexity of the interactions between the different components
of an extended open system.

The combination of Green's function  methods with a Density Functional Theory (DFT)
description of electronic states has become a standard approach to study charge
transport at the nanoscale~\cite{buon+01prb,tayl+01prb,bran+02prb,xue+02cp,calz+04prb}. 
However, care must be taken in comparing the computed transport characteristics to 
experiments~\cite{ever+04prb}. Indeed, some important features
--- such as electron correlations, dissipation, decoherence and
temperature effects --- are at the moment partly or fully neglected. These
deficiencies become more and more crucial when
the dimensions of at least one part of the system reach a confinement situation where
for instance electron-electron ({\it e-e}) interactions may become dominant.   
Recent observations of Kondo effect and Coulomb blockade in molecules connected 
to external electrodes~\cite{lian+02nat,park+02nat},
and in nanotubes with magnetic impurities~\cite{odom+00sci} or size 
confinement~\cite{jari+04nat},
indicate that correlations do play an important role in the mechanism of charge
transport in nano-devices. Whereas many efforts ~\cite{dive-pant00prb,koso03jcp,mait+03prb,
dela-gree04prl,ever+04prb}
have been directed to study other aspects of the transport problem, a standard
approach to include correlation effects does not yet exist. One may expect that some
of the difficulties in the interpretation of transport experiments on simple atomic
chains and individual molecules could be ascribed to the neglect or mistreatment 
of correlations.

In this Letter we develop a new method for the {\it ab initio} computation of 
quantum transport in the strong correlation regime, and then apply it to
specifically address the effect of {\it e-e} interactions on
electronic transport through atomic-scale conductors.
Following closely Meir and
Wingreen~\cite{meir-wing92prl}, we derive a Landauer-like expression for the current
through a correlated conductor between uncorrelated leads. Short range {\it e-e}
interactions in the conductor are described through the Three-Body Scattering (3BS)
formalism ~\cite{cala-mang94prb,mang+97prb}. The method
is implemented through use of ``maximally-localized'' Wannier functions~%
\cite{marz-vand97prb,calz+04prb} (MLWF).
A particularly well-adapted system for gauging the relevance of {\it e-e}
correlations is a late-transition-metal break-junction: in such a configuration, {\it
e-e} effects are negligible in the bulk but, as a consequence of dimensionality, may
acquire relevance in the junction region. 
We apply our method to simulate a model Platinum break-junction of varying
length. Our findings show a large suppression of the transmittance that we ascribe to
the inclusion of {\it e-e} elastic decoherence in the simulation. Moreover, the strong
conductance reduction, also for small wire lengths, suggests that correlation cannot
be neglected when studying transport properties of systems with localized electrons.

The system is modelled as three different regions, the left (L) and right (R)
electrodes and a central conductor (C).  We express our operators in a localized basis
set~\cite{datt95book} which allows us to write the Hamiltonian and the Green's
functions of the whole system as $3 \times 3$ block matrices defined on the basis in
the L, R, and C regions [$H_{xy}$, $G_{xy}(\omega)$ where $x,y$ = L,C,R].

The Hamiltonian reads:
%
%
\begin{equation}
\label{eq:MW_hamiltonian}
H = \sum_{l l' \in L {\rm or} R} H_{ll'} c_l^{\dagger} c_{l'} \, + \,
    H_{int} +
    \sum_{l \in L {\rm or} R \atop i \in C} \, \left[ H_{li} \, c_l^{\dagger} d_i +
           h.c. \right],
\end{equation}
where $c_l$ and $c_l^{\dagger}$ ($d_i$ and $d_i^{\dagger}$) are the one-electron
annihilation and creation operators in the leads (conductor). In the
above expression, the first term describes the L and R leads, $H_{int}$   the
conductor, the last term  the coupling of the conductor with the L and R leads. We
stress that the leads and the coupling Hamiltonian are non-interacting and all
the {\it e-e} interaction is restricted to the conductor.

From the continuity equation for the steady-state current in the
system and using the Keldysh non-equilibrium Green's function
formalism~\cite{haug-jauh96book_chapter4}, the following expression for the
current~\cite{meir-wing92prl} is derived:
%
%
\begin{equation}
\label{eq:MW_current}
 I = \frac{2ei}{h} \int \! \! d\omega \, \textrm{Tr} \left\{
         \left[ \Sigma_L^{<} - \Sigma_R^{<} \right]
         A_C
    + \left[ \Gamma_L - \Gamma_R \right] G_{C}^{<} \, \right\}.
\end{equation}
Here $A_C= i [G_C^r - G_C^a]$ is the spectral function and $G_C^{r,a,<,>}$ are the
(retarded, advanced, lesser, greater) Green's functions in the conductor. The trace
should be taken on the conductor. The interaction between conductor and leads is
described through the lead self-energies (SE's), defined as $\Sigma_{x}^{r,a,<,>}
(\omega) = H_{Cx} \, G_{x}^{r,a,<,>}(\omega) \, H_{xC}$, where $x$=L,R. Finally, the
$\Gamma_{L,R}$ terms in Eq.~(\ref{eq:MW_current}) are defined as twice the imaginary
part of the retarded lead-SE's, i.e. $\Gamma_{L,R} = i \left[ \Sigma_{L,R}^{r} -
\Sigma_{L,R}^{a} \right] $. Note also that $ \Sigma_{L,R}^{<} = i f_{L,R}\,
\Gamma_{L,R} $ and $ \Sigma_{L,R}^{>} = -i (1-f_{L,R})\, \Gamma_{L,R}$, where $f_L$
and $f_R$ are the Fermi occupation functions for the left and right leads.

While in the non-interacting case the above expression brings to the usual Landauer
formula~\cite{meir-wing92prl}, in the presence of interaction between electrons this
is no longer true and further assumptions are needed. We here adopt the {\it ansatz}
proposed by Ng~\cite{ng96prl} which relates $\Sigma^{<}_C$ ($\Sigma^{>}_C$) to
$\Sigma^{r,a}_C$, thus defining the statistics of energy levels,  
in the general out-of-equilibrium interacting case. The starting
point is the assumption~\cite{serg+02prb}:
%
%
\begin{equation}
   \label{eq:ansatz1}
   \Sigma^{<,>}_C(\omega) = \Sigma^{<,>}_{0 \, C}(\omega) \, \Lambda(\omega) \, ,
\end{equation}
where $\Sigma^{<,>}_{0 \, C} = \Sigma^{<,>}_{L} +
\Sigma^{<,>}_{R}$ refer to the non-interacting case and include
only the coupling with the leads, while the full $\Sigma^{<,>}_C$
include also {\it e-e} interactions. $\Lambda(\omega)$, which is called $A$ in 
other formulations~\cite{serg+02prb}, is
determined by the  identity $\Sigma^{>}_C - \Sigma^{<}_C =
\Sigma^{r}_C - \Sigma^{a}_C$ leading to:
%
%
\begin{equation}
   \label{eq:ansatzA}
   \Lambda(\omega) =  \left[ \Sigma^{r}_{0 \, C}(\omega) - \Sigma^{a}_{0 \, C}(\omega)
          \right]^{-1} \, \left[ \Sigma^{r}_C(\omega)-\Sigma^{a}_C(\omega) \right] \, .
\end{equation}
Here the interacting SE's take the form:
%
%
\begin{equation}
   \label{eq:total_SE}
   \Sigma^{r,a}_{C} (\omega) = \Sigma^{r,a}_L (\omega) + \Sigma^{r,a}_R (\omega) +
                           \Sigma^{r,a}_{corr} (\omega)
\end{equation}
where $\Sigma^{r,a}_{corr}$ account for the {\it e-e} interactions
(while $\Sigma^{r,a}_{0 \,C}$ just drop this last term).
Relations~(\ref{eq:ansatz1}-\ref{eq:total_SE}) are the  key to relate
Eq.~(\ref{eq:MW_current}) to a Landauer-like formula.
In fact, following Eq.~(6) of
Ref.~[\onlinecite{serg+02prb}] and using  $ G_{C}^{<} = G_{C}^{r} \Sigma^{<}_{C}
G_{C}^{a}$, it is possible to derive $ G_{C}^{<}  = i \, G_{C}^{r} \, [ f_L \Gamma_L +
f_R \Gamma_R ] \, \Lambda G_{C}^{a} $ and therefore \mbox{$ G_{C}^{r} - G_{C}^{a} =
-i\, G_{C}^{r} [ \Gamma_L + \Gamma_R ] \,
 \Lambda G_{C}^{a}$}.
These steps lead to the final expression for the current:
%
%
\begin{equation}
   \label{eq:landauer_gen_current}
   I = \frac{e}{\hbar} \int \, \frac{d\omega}{2\pi} \,
       \left[ f_L - f_R \right] \,
       \textrm{Tr} \left\{ \,\,  \Gamma_L \, G_{C}^{r} \,
       \Gamma_R \, \Lambda \,
       G_{C}^{a} \,\, \right\}. 
\end{equation}
We remark that the {\it e-e} correlation plays a twofold role: it renormalizes the
Green's functions, which should now be calculated for the interacting system, and also
modifies the expression for the current, introducing the corrective factor
$\Lambda(\omega)$. The quantity traced in Eq.~(\ref{eq:landauer_gen_current}), even if
not a true transmittance across the scattering region due to the breakdown of
Landauer's theory, still plays the same role as regards transport. For this reason we
refer to it as an {\it effective transmittance}  and compare it to the transmittance
of the non-interacting case. In particular, since the imaginary part of the {\it e-e}
self-energy exactly vanishes at the Fermi energy ($E_F$), the correction
$\Lambda(\omega=E_F)$ becomes the identity operator and the effective transmittance
computed at $E_F$ is proportional to the conductance at zero temperature, provided
that the SE's are calculated for the interacting system.

An accurate evaluation of the effective transmittance needs to be based on: (i) a good
description of the non-interacting system, and (ii) the calculation of the {\it e-e}
correlated SE to include many-body effects arising from the interaction in the
conductor. The first problem is solved by exploiting a recent methodology for the {\it
ab initio} calculation of the transmittance in the coherent transport
regime~\cite{calz+04prb}. The ground state of the mean-field system is described
within the Local Density Approximation to DFT, using norm-conserving pseudopotentials
and a plane-wave  basis set~\cite{pwscf}. The geometry of the L-C-R nano-junction are
as in Ref.~\cite{buon+01prb}. To obtain a consistent description of the system in a
real-space localized basis set, the MLWF's are computed both for the leads and for the
conductor~\cite{want}. The details of this transformation and its application to 
electronic transport are described elsewhere~\cite{marz-vand97prb,calz+04prb}. 
We remark that the basis change from
Bloch eigenvectors to MLWF's preserves orthonormality and completeness in the original
Hilbert subspace, thereby avoiding typical problems arising very often for other
localized basis functions. The same features allow us to use a minimal basis set for
computing transport properties, while employing the system independent plane-wave
basis set for the DFT calculation.

To discuss the inclusion of {\it e-e} correlations we first need to define the
interaction hamiltonian in Eq.~(\ref{eq:MW_hamiltonian}). We focus on the short range {\it
e-e} interaction for two main reasons. On one hand, this regime is characterized by
strong deviations from the non-interacting behavior, for instance in terms of
quasiparticle lifetimes (which include elastic decoherence in the transport
formalism). On the other hand, it allows us to adopt an Anderson-like
form~\cite{maha81book} of the interaction which is suitable for the localization in
the conductor region only, as required by our approach.

In this work the {\it e-e} self-energy $\Sigma^{r}_{corr}$ is computed using a
non-perturbative approach based on an effective Anderson hamiltonian, whose $U$
Coulomb integrals could be either calculated {\it ab initio}~\cite{spri-arya98prb} or
used as adjustable input parameters. It is solved by means of a configuration
interaction scheme with up to three bodies (3BS) added to the non-interacting Fermi
sea: two (one) electron(s) and one (two) hole(s) for conduction (valence) states. This
method has been successfully applied to describe photoemission experiments on strongly
correlated systems~\cite{mang+97prb}. The 3BS self-energy is formally
given~\cite{mang+97prb} as a sum over projectors onto atomic states (those with
non-negligible $U$ integrals) and thus can be properly localized in the conductor.


We now come to the model system: a Pt atomic wire of varying length, where
the correlation is switched-on only on a finite number $N_C$ of atoms.
It is worth noting that late $5d$ transition-metal atomic wires have been demonstrated in
break-junction experiments~\cite{agra+03prep}. Moreover, in the case of Pt, correlation
effects are expected to strongly increase their importance in passing from the bulk
leads to the confined wire as an effect of the reduced dimensionality. 
Therefore, our model system, while neglecting the full complexity of the wire-lead 
interfaces, retains the basic geometric construct of a correlated wire between 
uncorrelated leads in a Pt break-junction. Hence, it can be considered suitable to
emphasize the effects of correlation and decoherence in nano-junctions, whereas the
effects of dimensionality at the junctions (3D instead of 1D leads) is already
known~\cite{agra+03prep}.

\begin{figure}
   \includegraphics[clip,width=0.48\textwidth]{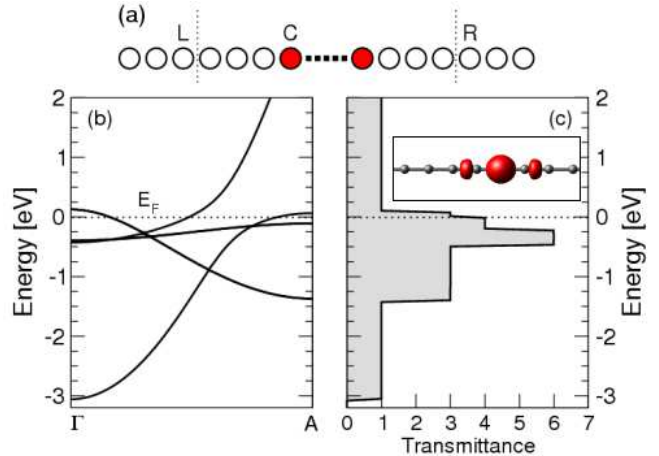}
   \caption{\label{figure1} (color online).
            (a) System layout divided in the leads and conductor regions.
                Open (full) circles are uncorrelated (correlated) Pt atoms.
            (b) Band structure of the Pt infinite chain.
                $A$ is the edge of the one-dimensional Brillouin zone.
            (c) Transmittance of the non-interacting conductor region.
                The inset shows the isosurface plot of a computed MLWF.
                The Fermi energy is set to zero.
   }
\end{figure}

The setup is described in Fig.~\ref{figure1}(a). For region C we
consider a supercell containing 11 atoms with an interatomic
separation of 3.0 \AA{}, reasonable for experiments under stretching (such as
break-junctions~\cite{agra+03prep}). The same geometry is adopted
for the leads, which are modelled as semi-infinite wires treated
at the single-particle level. Since the $sd$-bands of the Pt
chain~[Fig.~\ref{figure1}(b)] form an isolated
subset~\cite{marz-vand97prb}, we can extract a manifold of MLWF's
which span the same subspace.  As an example, a particular wave
function well localized within two bond-lengths is shown in the
inset of Fig.~\ref{figure1}(c). This MLWF subspace allows us to
reproduce with good accuracy states more than 2 $eV$ above the
Fermi energy, enough to describe transport properties in this
system where $p$-orbitals are expected to play a negligible role.
Figure~\ref{figure1}(c) shows the computed transmittance for the
non-interacting Pt chain, which counts, for each energy, the
number of channels (bands) available for charge transport, leading
to a zero temperature conductance of $3 G_0$ ($G_0 = 2e^2/h$).

\begin{figure}
   \includegraphics[clip,width=0.47\textwidth]{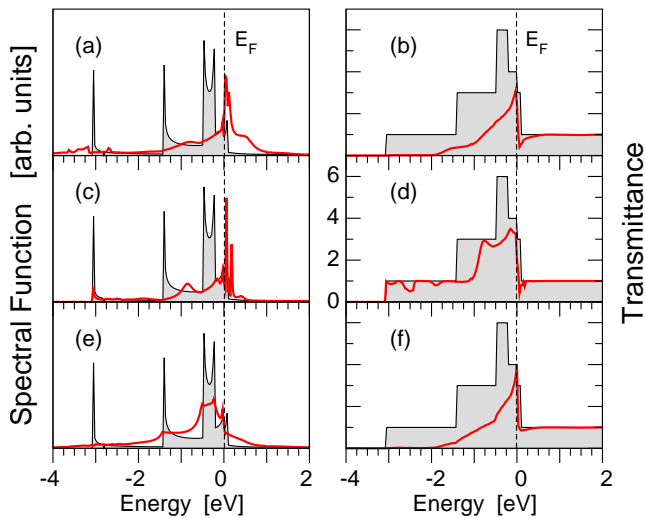}
   \caption{\label{figure2} (color online).
        Curves: Spectral function projected on the interacting region (left)
        and effective transmittance (right) for the
        case of three correlated atoms in the Pt chain.
        Shaded areas: mean-field reference results.
        $E_F$ is set to zero.
        (a) and (b) show the computational results obtained with the
        total correlation SE $\Sigma_{corr} = \Sigma_{3BS}$.
        The (c,d) [e,f] plots are obtained using only the
        hermitian [antihermitian] SE $\Sigma_H$ [$\Sigma_A$].
   }
\end{figure}

In Fig.~\ref{figure2}(a,b) we report the spectral function
projected on the interacting region and the effective
transmittance for a chain with three correlated atoms (U = 2.0
$eV$). The many-body spectral function shows a splitting of the
$d$-bands and a slight upward energy shift, which is consistent
with our short range interaction picture based on the physics of
the Anderson hamiltonian. The transmittance is strongly suppressed
by the inclusion of correlation [Fig.~\ref{figure2}(b)], which is
particularly effective in the hole region~\cite{landauer_vs_gen}.
This can be understood considering that on-site correlation arises
from the strongly localized Pt $d$-orbitals, that are largely
occupied and produce major features below $E_F$.

The introduction of {\it e-e} interactions leads to quasi
particles characterized by energy and lifetime (finite broadening
of their spectral features) formally accounted for by the
hermitian (\emph{H}) and the antihermitian (\emph{A}) parts of the
SE operator. Note that in more common correlation treatments, such as 
LDA+U, the SE in hermitean and lifetimes are consequently neglected.  
Although these components must obey analytic
constraints~\cite{mang+97prb}, we analyze them separately to
highlight their very effect on transport. First of all we divide
the SE operator in \emph{H} and \emph{A} contributions,
$\Sigma_{corr} = \Sigma_H + \Sigma_A$ where:
\begin{equation}
   \label{eq:sigma_HA}
   \Sigma_{H,A} (\omega) = \frac{1}{2} \left[ \Sigma (\omega) \pm
\Sigma^{\dagger}
                           (\omega)\right] \, .
\end{equation}
In Fig.~\ref{figure2}(c,d) and (e,f) we show our results when
using $\Sigma_H$ or $\Sigma_A$, respectively, instead of the full
correlation SE from 3BS. By definition $\Sigma_{A}$ vanishes at
$E_F$, thus the only contribution to the zero temperature
conductance comes from the \emph{H} part of $\Sigma$. Moreover, at
energies different from $E_F$ the major quenching factor on the
transmittance is due to the \emph{A} part of $\Sigma$, while its
effect on the spectral function is just a slight broadening of the
main peaks. Note that the decrease of the effective transmittance
due to $\Sigma_H$ is related to energy misalignment of channels
between the conductor and the leads, while that due to $\Sigma_A$
accounts for {\it e-e} scattering. Our results thus indicate that
{\it e-e} scattering plays a fundamental role in suppressing
effective transmittance for systems with strong short range
correlation. This effect should be hidden in the low temperature
conductance but evident in the $I-V$ characteristics
[Eq.~(\ref{eq:landauer_gen_current})]: a suggestion for further
experimental analysis.

\begin{figure}
   \includegraphics[clip,width=0.47\textwidth]{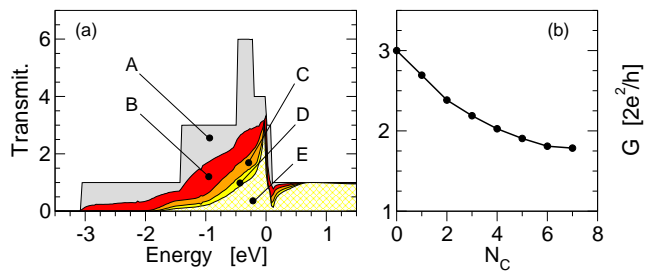}
   \caption{\label{figure3} (color online).
           (a) Effective transmittance against energy for variable
               number of correlated atoms in the chain:
               $A$ is the reference mean-field bulk transmittance.
               $N_C$=1,3,5,7 in B,C,D,E.
           (b) Conductance as function of the number of
               correlated atoms in the Pt chain.
               Saturation is observed between $N_C$=6 and 7.
   }
\end{figure}

Fig.~\ref{figure3} reports the computed conductance and effective transmittance for
variable number of correlated atoms $N_C$. Figure~\ref{figure3}(a) shows the effective
transmittance curves for various $N_C$ values and demonstrates that the effective
transmittance decreases with increasing $N_C$. This behavior is consistent with the
presence of a scattering mechanism, according to which one expects a vanishing
effective transmittance at $\omega\neq E_F$ in the limit of infinitely long correlated
wires. Figure~\ref{figure3}(b) reveals also a rapid decrease of the conductance at
small values of $N_C$. The last result shows firstly that the conductance is
renormalized as well as the transmittance, although the imaginary part of the SE
vanishes at $E_F$. Secondly, it suggests that the effect of on-site correlation should
be experimentally measurable also for short atomic chains (such as those produced by
break-junctions).

In conclusion, we derived a generalized Landauer-like expression for the current
[Eq.~(\ref{eq:landauer_gen_current})] and the conductance in the presence of {\it e-e}
interactions. The formalism is suitable for a fully {\it ab initio} implementation
that we realized using a basis set of maximally-localized Wannier functions for the
Green's functions and the 3BS formalism for the {\it e-e} self-energy. We applied the
method to a finite Pt wire and found  a renormalization of the conductance and a
strong quenching of the transmittance as a consequence of the {\it e-e} scattering.
Our results suggest that the inclusion of electron correlation for systems with strong
short-range interactions is essential for an accurate description of current and
conductance.

We acknowledge illuminating discussions with C. A. Rozzi and V. Bellini. Funding was
provided by the EC through project IST-2001-38951 and TMR network ``Exciting'', by
INFM through ``Commissione Calcolo Parallelo'' and by MIUR (Italy) through
``FIRB-NOMADE''. M.B.N. and A.C. acknowledge the Petroleum Research Fund of the ACS.
M.B.N. also acknowledges the Mathematical, Information and Computational Sciences
Division, Office of Advanced Scientific Computing Research of the U.S. DOE under
contract No. DE-AC05-00OR22725 with UT-Battelle. MJC acknowledges support from INFM
$S^3$, Italy, FAPESP and CNPq, Brazil.


\begin{thebibliography}{10}
\bibitem{landauer_vs_gen}
Note that these results are mainly due to the renormalization of the GF's in
  Eq.~(\ref{eq:landauer_gen_current}) while the correction term
  $\Lambda(\omega)$ plays a minor role. This finding may be caused by the very
  one-dimensional nature of the considered system and further investigation is
  needed.
\end{thebibliography}
\end{document}